# Effect of the isoelectronic substitution of Sb for As on the magnetic and structural properties of LaFe(As$_{1-x}$Sb$_x$)O

S.J.E. Carlsson[*], F. Levy-Bertrand[*], C. Marcenat[†], A. Sulpice[*], J. Marcus[*], S. Pairis[*],
T. Klein[*], M. Núñez-Regueiro[*], G. Garbarino[‡], T. Hansen[§], V. Nassif[*,§], P. Toulemonde[*]

[*] *Institut Néel, CNRS and UJF 25, 38042 Grenoble, France*
[†] *SPSMS, UMR-E 9001, CEA-INAC/UJF, 38054 Grenoble, France*
[‡] *ESRF, 38043 Grenoble, France*
[§] *Institut Laue et Langevin, 38042 Grenoble, France*

**Abstract:**

The antiferromagnetic order and structural distortion in the LaFe(As$_{1-x}$Sb$_x$)O system have been investigated by powder neutron diffraction and physical properties measurements. Polycrystalline samples of LaFe(As$_{1-x}$Sb$_x$)O (x<0.5) were prepared using solid state synthesis at ambient and high pressure. We find that the isoelectronic substitution of Sb for As decreases the structural and magnetic transition temperatures but, contrary to the effects of phosphorus substitution, superconductivity is not induced. Instead a slight increase in the Fe magnetic moment is observed.

## I. INTRODUCTION

Since the discovery of high $T_C$ in the new oxy-pnictide superconductors[1] there has been an extensive amount of studies focused on understanding the underlying mechanism of the superconductivity in this family of compounds and to raise the $T_C$ further.[2] The characteristics of these materials are similar to those of the copper-based superconductors such as the layered structure, the loss of the antiferromagnetic (AFM) ground state and subsequent emergence of superconductivity with chemical doping.[3,4] However, in contrast to the cuprates, superconductivity can be induced in certain pnictide compounds not only by charge doping but also by applying external pressure[5,6,7] or by isoelectronic doping[8,9,10,11]. In the iron-arsenides, the superconductivity appears to be unconventional[12,13] and closely related to the magnetism.[14,15,16,17] It is known that the Fermi surface can be modified by changes in the crystal structure parameters[18,19] but the complex relationship between the superconductivity, the magnetic order and the structural parameters are not yet fully understood. The shape of the FeAs$_4$ tetrahedron seems to play a decisive role with, in particular, the Fe-As bond distance having a strong influence on the Fe magnetic moments according to density-functional theory (DFT) calculations.[11,20,21]

LaFeAsO, one of the parent compounds of the iron-arsenide superconductors, undergoes a structural phase transition from a tetragonal to an orthorhombic structure at $T_S$~ 160 K.[22] A spin-density wave (SDW) type ground state with long range antiferromagnetic order is formed below the magnetic transition temperature ($T_N$) ~137 K. Partial substitution of isovalent P for As leads to a suppression of the SDW order and the emergence of superconductivity. The $T_C$ reaches ~10 K for an optimal doping level of 25-30%.[9] A similar behaviour is observed in BaFe$_2$(As$_{1-x}$P$_x$)$_2$ where the maximum $T_C$ is reached for x=0.26.[8,23] This is thought to be a result of the shrinking of the Fe-As/P bond caused by the introduction of the smaller P atoms into the arsenide system. In fact, the isoelectronic substitution of phosphorus for arsenic, i.e. without changing the carrier density, is an ideal tuning parameter for investigating the boundary between magnetic order and superconductivity. However, it is also possible to isoelectronically tune the system by replacing arsenic with antimony. Considering that Sb is larger than As, the effect on the structural parameters should be very different from that of P substitution and thus also the electronic behaviour. So far, there have been very few studies on such systems. However, based on DFT calculations, it was suggested by Moon *et al.* that LaFeSbO is a good candidate for higher $T_C$ when considering the spin-fluctuation-mediated superconducting scenario[24]. Their results showed that in the SDW state, the Fe spin moment in LaFeSbO is larger than that of LaFeAsO and LaFePO. Furthermore, the density of states (DOS) at the Fermi energy (*Ef)* is almost doubled indicating a stronger electron-spin-fluctuation coupling in LaFeSbO which is necessary for higher $T_C$. This theory appears to be supported by two previous experimental studies of the partial substitution of Sb for As which show an enhancement of the $T_C$ from 28.5 K to 30.1 K in fluorine doped LaO$_{0.8}$F$_{0.2}$FeAs$_{1-x}$Sb$_x$ samples with x=0.05.[25,26] Increased Sb content (x=0.2 for LaO$_{0.1}$F$_{0.1}$FeAs$_{1-x}$Sb$_x$) suppresses the superconductivity and the SDW order is recovered.[26] These experimental reports focus on samples doped with fluorine and the structural and magnetic properties of the isoelectronically substituted LaFeAs$_{1-x}$Sb$_x$O compounds have not been studied in detail.

Therefore, we have explored the effect of Sb doping on the crystal and magnetic phase transitions of LaFeAsO using neutron powder diffraction along with magnetisation, transport and specific heat measurements. In agreement with the theoretical predictions, we find a slight increase of the Fe magnetic moment in the AFM phase. Moreover, we find that although the addition of Sb decreases the temperatures associated with the structural phase transition ($T_S$) and AFM ordering ($T_N$) it cannot by itself induce superconductivity in LaFeAsO.

## II. EXPERIMENTAL DETAILS

Polycrystalline samples of LaFeAs$_{1-x}$Sb$_x$O were prepared by two different synthesis routes. The first method was a two-step solid-state reaction similar to that reported by Zhu *et al.* using FeAs, FeSb, La and dried La$_2$O$_3$ as starting materials.[27] The FeSb was obtained by reacting Fe and Sb powder at 700°C for 96h in evacuated silica tubes. A mixture of the reactants was ground and pressed into pellets in a glove box filled with Ar, then sealed in evacuated silica tubes and heated at 800°C for 12h immediately followed by a second plateau at 1100°C for 48h. Secondly, samples were synthesised under high pressure and high temperature (HP-HT). Small pellets of the same starting mixture as above were placed in a BN crucible under an inert argon atmosphere and subjected to a pressure of ~3.6 GPa and 1100°C for 1h using a Belt-type apparatus. The phase purity of the resulting powders was examined by powder X-ray powder diffraction using Cu Kα radiation. Analysis of the chemical composition was carried out on sintered pellets of the compounds by field emission scanning electron microscopy (FESEM) on a ZEISS Ultra+ microscope equipped with a Bruker EDX system.

Neutron powder diffraction was used to study the structural and magnetic phase transitions in LaFeAs$_{1-x}$Sb$_x$O with x=0, 0.2 and 0.4. Initially, an experiment was made on the D1B instrument at Institut Laue Langevin (ILL, Grenoble) using a monochromatic beam at λ=2.52 Å. During this study, the AFM transition on cooling was localised. In order to study the structural transition (from tetragonal to orthorhombic lattice) and the magnetic ordering, a second experiment was then performed on the high-flux diffractometer D20 at the ILL in the temperature range 2-300K. The instrument was used in the high-resolution mode with a wavelength of λ=1.87 Å (from the Ge(115) reflection of the monochromator). This made it possible to observe the magnetic contributions to the diffraction pattern as well as the splitting of the nuclear reflections which is the signature of the structural phase transition. All the diffraction data were analysed by the Rietveld method[28] using the GSAS[29] suite of programs.

The DC electrical resistivity of the samples was measured by a standard four-point technique in the temperature range 4-300 K using silver paste for the electrical contacts. Magnetisation measurements were performed between 2 and 280 K in a static magnetic field of 1 T with a SQUID magnetometer (Quantum Design). The temperature dependence of the specific heat in zero magnetic field was investigated using an AC microcalorimetry technique together with a diode heater and a thermocouple to record the temperature oscillations.

## III. RESULTS

### A. Structural phase transition probed by x-ray and neutron powder diffraction

Rietveld analysis of the x-ray and neutron diffraction data show that the samples consist mainly of the "LaFeAsO type" phase with small amounts of the impurities FeSb, FeAs and $La_2O_3$ (transformed into $La(OH)_3$ when exposed to air). Figure 1(a) shows the result from the refinement of the x-ray diffraction pattern of $LaFeAs_{0.6}Sb_{0.4}O$ at 300 K. The variation of the lattice parameters (obtained from the x-ray and neutron diffraction patterns) with the nominal Sb content are presented in Fig. 2. There is a clear increase in the $a$ and $c$ lattice parameters of both types of samples as As is replaced by the larger Sb, confirming that the latter is being incorporated into the LaFeAsO structure. A deviation from the continuous increase of the lattice parameters was observed above 40% Sb indicating that the limit for the amount of Sb in the solid solution has been reached. EDX analysis confirmed that the highest Sb content achieved was 38(2)at% using HP synthesis and 39(2)at% for AP synthesis. The lattice parameters of the HP-HT samples are smaller than those made at ambient pressure (AP). This is most likely an effect of the high pressure synthesis and the presence of oxygen vacancies in the LaFeAsO phase. Extrapolation of the lattice parameters of the AP samples up to pure LaFeSbO (Fig. 2) gives a value of $a$, which is in good agreement with that estimated by Lébegue et al.[30] but lower than that from Moon et al.[24] (based on DFT calculations). The $c$ parameter is slightly smaller than predicted by both studies.

In this paper we mainly discuss the properties of our sealed tube samples made at ambient pressure in order to avoid any effects created by oxygen vacancies probably present in the HP samples. Furthermore, the sample volume is larger which is necessary for the neutron diffraction experiments and comparisons can be made with previous studies of the LaFeAsO system where similar AP synthesis methods are most often used.

Upon cooling, LaFeAsO undergoes a structural phase transition from tetragonal (P4/*nmm*) to orthorhombic (C*mme*) at a temperature, $T_S$, of ~160 K.[22,31,32] In order to determine the effects of Sb substitution on the structural transition in $LaFeAs_{1-x}Sb_xO$ we used neutron diffraction at low temperature to study the structural details of two samples with x=0.2 and x=0.4 made by ambient pressure synthesis. Figure 1(b) shows the results of the Rietveld refinement of the $LaFeAs_{0.6}Sb_{0.4}$ diffraction pattern at 2 K. The structural parameters obtained from the Rietveld analysis are presented in table I. For the un-doped LaFeAsO, the parameters were extracted from the literature.[32] The oxygen content of the samples was examined by varying the oxygen occupancy in the refinement of the 300 K neutron diffraction patterns. Within error the values obtained were equal to one showing that the compounds are fully oxidised. Thus, the occupancy was fixed at unity for the final refinements. The Sb content could not be determined from the neutron data as the coherent scattering amplitudes for As and Sb are too similar. However, the Sb content based on the Rietveld analysis of the x-ray diffraction data was 22(2) at% and 39(2)at% for the x=0.2 and x=0.4, respectively. In addition, the corresponding values obtained from EDX analysis were 19.6(9)at% and 40.5(9)at% which are very close to the nominal compositions. Therefore, the nominal values of the Sb content (kept fixed with temperature) were used in the neutron diffraction refinements. Minor impurity phases of FeSb (4.2% for x=0.2 and 5.3% for x=0.4) and $La_2O_3$ (6.1% for x=0.2 and 6.2% for x=0.4) were included in the refinements. A good quality of the fits was achieved using the structural models for LaFeAsO (listed above) as represented by the low residual values listed in table I.

Figure 3(a-b) shows the neutron diffraction profiles of the $(220)_T$ reflection (tetragonal cell) at various temperatures for the samples of x=0.2 and x=0.4. A split into the $(400)_O$ and $(040)_O$ reflections (orthorhombic cell), revealing the structural phase transition, is clearly visible in both sets of data. It is difficult to determine at what temperature to switch between the *Cmme* and P4/*nmm* models but based on the criteria applied in previous studies of LaFeAsO (ref. 31,32), the *Cmme* space group was used to refine the Ln-1111 structure when the difference between the refined values of *a* and *b* was greater than the error associated with these parameters. In our case this approach was applied to data collected below 200 K. Comparing the two profiles, it appears that the structural transition begins at a higher temperature in the x=0.2 sample. Figure 4(a) shows the temperature dependence of the lattice parameters. As seen in pure LaFeAsO[31], there is a rapid convergence of the parameters *a* and *b* of $LaFeAs_{0.8}Sb_{0.2}O$ up to 150 K on increasing temperature. The trend is more difficult to distinguish in the sample of x=0.4 due the limited sets of temperature points collected but the lattice parameters appear to converge at a lower temperature. This is confirmed when looking at the temperature dependence of the orthorhombicity $(a-b)/(a+b)$ shown in the inset of Fig. 4(b). The data for LaFeAsO are taken from ref. 32. Considering the changes of the slope of the orthorhombicity curve, $T_S$ can be determined to ~150 K for $LaFeAs_{0.8}Sb_{0.2}O$ and ~130 K for $LaFeAs_{0.6}Sb_{0.4}O$. It can also be seen that the orthorhombic distortion (extrapolated at 0 K) becomes larger with increased Sb substitution.

## B. Magnetic phase transition probed by neutron powder diffraction

Figure 3(c) shows the low angle region of the neutron powder diffraction patterns of LaFeAs$_{0.6}$Sb$_{0.4}$O collected at D20 on decreasing temperature. As observed in the neutron powder diffraction profiles from the D1B experiment (not shown), additional peaks appear below 120K. The same reflections were also observed in the LaFeAs$_{0.8}$Sb$_{0.2}$O sample below 140 K. These are the (1,0,3/2) and (1,2,1/2) magnetic Bragg contributions, as seen in LaFeAsO, indexed in the magnetic unit cell $a_N \times b_N \times 2c_N$ (where $a_N$, $b_N$ and $c_N$ are the nuclear lattice parameters) with a stripe-type AFM arrangement of the Fe moments.[22,32] The data were refined using a nuclear phase and a purely magnetic phase, both with the orthorhombic space group *Cmme* and a scale factor of 1:2 (nuclear:magnetic) due to the doubling of the *c*-axis in the magnetic unit cell. Because of the small number of weak magnetic peaks, it was not possible to determine the alignment of the magnetic moments along the *a* or *b* axis. The spin direction was therefore fixed along *a* according to the model suggested by Qureshi *et al*.[32] This gave a good fit to the data with a resulting Fe magnetic moment of 0.95(10)$\mu_B$ and 0.98(9) $\mu_B$ for the x=0.2 and x=0.4 samples, respectively, at 2 K. These are in agreement with the Fe moments extracted from the Rietveld refinements of the D1B data (not shown) which were 0.96(18)$\mu_B$, 1.02(13)$\mu_B$ and 1.20(17)$\mu_B$ for x=0, 0.2 and 0.4, respectively. However, our values are larger than those first reported for pure LaFeAsO of ~0.36$\mu_B$ [22, 31] but nearer to the recently reported ~0.63(1)$\mu_B$[32] and ~0.8$\mu_B$ (single crystal).[33] Furthermore, the moments obtained in this study are comparable to those of the "122" compounds (AFe$_2$As$_2$ with A=Ca, Sr and Ba) which are in the range of 0.8-1.1$\mu_B$. This is in line with the suggestion of an effective S≈½ on the iron sites.[33,34] The slight increase of the Fe moment with Sb substitution is in agreement with the DFT calculations of Moon *et al*.[24] who predicted that $\mu_B$ would be 10% higher in the hypothetical LaFeSbO than in LaFeAsO. The temperature variation of the magnetic moment is presented in Fig. 4(b). Both samples show similar behaviour but $\mu_B$(T) starts to drop at a lower temperature for x=0.4. $T_N$ can be estimated to ~120 K and ~140 K for LaFeAs$_{0.6}$Sb$_{0.4}$O and LaFeAs$_{0.8}$Sb$_{0.2}$O, respectively. It can also be seen that the variation of the magnetic moment with temperature follows the changes in the orthorhombicity (inset Fig. 4(b).) implying that the magnetic and structural transitions are closely related. In addition, this figure suggests that a larger orthorhombic distortion strengthens the magnetic interactions. The relationship between the structural distortions and the magnetic properties will be discussed in the following section.

## C. Physical properties measurements

The temperature dependence of $c_p$, $M/H$ and $\rho$ of the LaFeAs$_{1-x}$Sb$_x$O (x=0, 0.2, 0.4) samples used in the neutron diffraction study are summarised in Fig. 5. All quantities have been normalised for comparison. The structural and magnetic phase transitions are accompanied by anomalies in the data of all three measurements.

Figure 5(a) illustrates the anomalous contributions to the specific heat, $\Delta c_p/T$. To highlight the effect of the transitions, a background determined by a polynomial function fitted above and below the transition region was subtracted from the measured data (shown in the inset Fig. 5(a) for x=0.2). The broad $c_P$ anomaly can be modelled as two overlapping peaks corresponding to $T_S$ and $T_N$ for the three compositions. There is a clear shift of the position of the anomalies towards lower temperatures with increased Sb content i.e. a decrease in the transition temperatures just as observed in the neutron diffraction data. The temperature difference between the structural and magnetic transitions is smaller at higher doping levels. Based on the method of Klauss et al.[35] and Kondrat et al.[36], $T_S$ and $T_N$ can be determined as 163(2) K and 144(2) K, respectively, for our LaFeAsO which is similar to previously reported values.[31,36,37] In LaFeAs$_{0.6}$Sb$_{0.4}$O (LaFeAs$_{0.8}$Sb$_{0.2}$O), $T_S$ and $T_N$ have been reduced to 130(2) K (153(2) K) and 116(2) K (135(2) K), respectively. This is in good agreement with Wang et al.[26] who found that $T_S$=144 K and $T_N$=133 K for LaFeAs$_{0.67}$Sb$_{0.33}$O.[26]

The impact of the Sb substitution on the phase transitions can also be seen in the magnetisation measurements presented in Fig. 5(b). A smooth background generated by a Curie-Weiss fit was removed from the LaFeAs$_{0.6}$Sb$_{0.4}$O data in order to remove the magnetic contributions from the FeSb impurity and make the anomaly clearer. At the onset of the structural phase transition there is a decrease in $M/H$ caused by the strengthening of the antiferromagnetic interactions. The inset of Fig. 5(b) shows the derivative $d(M/H)/dT$, where the phase transitions are visible as two overlapping peaks with a temperature shift between the samples of x=0.2 ($T_S$=152(2) K, $T_N$=135(2) K) and x=0.4 ($T_S$=128(2) K, $T_N$=116(2) K) as observed in the $c_P$ data.

Figure 5(c) illustrates the temperature dependence of the resistivity $\rho(T)$ for LaFeAs$_{1-x}$Sb$_x$O, x=0, 0.2 and 0.4. The compounds show similar behaviour and they all exhibit an anomaly with a local maximum in $\rho(T)$ associated with the structural phase transition as previously reported for LaFeAsO.[31,35,36] No superconductivity was observed down to 4 K. The derivative of the resistivity data is displayed in the inset of Fig.5(c). A sharp increase in $d\rho/dT$ is observed at the structural phase transition and the maximum value is close to the magnetic transition temperature. Due to the polycrystalline nature of the samples and their possible local inhomogeneity (uneven Sb distribution in the grains of the La-1111 phase) it is difficult to make a comparison of the actual values of the resistivity. However, the anomaly of $\rho(T)$ originating from the SDW transition shifts to lower temperatures with increased Sb substitution. It also becomes less pronounced as the drop in the resistivity below $T_S$ is reduced and the gap between $T_S$ and $T_N$ diminishes as seen in the specific heat and the magnetisation data.

The structural and magnetic transition temperatures derived from the physical properties measurements following the method of Klauss *et al.*[35] and Kondrat *et al.*[36] are summarised in Fig. 6(a) together with corresponding values determined from the neutron diffraction data. It clearly illustrates that Sb substitution leads to a decrease in both transition temperatures and also a reduction of the temperature difference between $T_S$ and $T_N$. The results from the specific heat and the magnetisation measurements are in good agreement with the structural analysis whereas the resistivity data gives comparatively slightly lower values. An explanation may be that the resistivity is not a measurement of the bulk properties and therefore it is more sensitive to any inhomogeneities of the samples, as mentioned previously. Extrapolation of $T_N$ to higher Sb content based on the data in Fig. 6(a) gives a $T_N \sim 60(5)$ K for x=1.0 suggesting that the magnetic order is present even in pure LaFeSbO. This supports the predictions from DFT calculations that LaFeSbO will have a SDW-type AFM ground state.[24]

## IV. DISCUSSION

Our results show that the isoelectronic substitution of Sb and P for As in LaFeAsO has very different effects on the electronic and magnetic properties. Although a decrease in $T_S$ and $T_N$ is observed in both cases, contrary to the case of phosphorous, the addition of antimony alone can not totally suppress the SDW order nor induce superconductivity. To understand these differences a closer investigation of the structural properties is necessary. The variation of some selected structural parameters of LaFeAsO with P (taken from ref. 9) and Sb content are presented in Fig. 6(b). It is clear that the structural distortions induced by antimony are exactly opposite to those caused by phosphorous. Studies of the $BaFe_2(As_{1-x}P_x)_2$ and $CeFeAs_{1-x}P_xO$ systems suggests that the SDW transition and the magnetic moment are very sensitive to the iron-pnictide distance.[11,38,39] DFT calculations also reveal a clear dependence of $\mu_{Fe}$ with the FeAs bond distance. Our structural refinements show that Sb substitution for As leads to a substantial increase in the Fe-Pn distance, 5.3% between x=0 and x=0.4, contrary to a 1.2% decrease for LaFeAs$_{1-x}$P$_x$O, x=0.3 (ref. 9). This could explain why we observe a small increase in the magnetic moment and why the AFM transition is not suppressed in LaFeAs$_{1-x}$Sb$_x$O. The decrease of the transition temperatures with increased Sb content may be a result of a weakening in the interlayer magnetic interactions caused by the expansion of the *c*-axis which in turn increases the distance between the iron layers. Furthermore, introduction of Sb can lead to disorder in the FeAs layer and hence a weakening of the magnetic interactions. Comparatively, the expansion of the Fe-Fe bond is much smaller, around 1.2% between the samples of x=0 and x=0.4. This is consistent with the 2% increase observed when going from LaFePO to LaFeAsO and confirms that the in-plane Fe-Fe distance is not the critical factor in the stabilisation of the AFM state.[40] These findings also indicate that the impact of antimony on the SDW order in LaFeAsO is much smaller than the effects induced by the equivalent amount of phosphorous. While 30% P is enough to totally quench the magnetic moment and suppress the SDW order[9], 40% Sb only leads to a decrease in $T_S$ and $T_N$ and a slight increase of the magnetic moment. Hence, increasing the Fe-As distance has a smaller influence on the magnetic properties than reducing it.

In the literature there are also several examples of electron doping, such as the substitution of Ni (ref. 41), Pt (ref. 42), Mn (ref. 43), Cr (ref. 44,45) and Co (Ref. 46) directly at the Fe sites in the 122 compounds. Amongst them, effects on the structural and magnetic properties similar to those observed in our LaFeAs$_{1-x}$Sb$_x$O samples have been reported. A decrease in $T_N$ coupled with strong magnetism and the absence of superconductivity have been observed for hole doping in Sr(Fe$_{1-x}$Mn$_x$)$_2$As$_2$ (ref. 43) and Ba(Fe$_{1-x}$Cr$_x$)$_2$.[44,45] Comparing Sb substitution with Mn or Cr doping, in terms of structural modification, the same changes are induced in both cases. Increased dopant concentration leads to an increase in the Fe-As distance and the As-Fe-As angles go towards the ideal value (see table I). At high doping levels, ~40%, of Mn and Cr the AFM transition is totally suppressed and a competing magnetic ground state different from the superconducting one emerges. This is not observed in our LaFeAs$_{1-x}$Sb$_x$O series. However, the impact of Sb can be expected to be less significant since it does not have a magnetic moment and the substitution is not directly on the magnetic Fe site. First principles calculations on the Mn doping effect in SrFe$_2$As$_2$ show that for hole doping, the Fe-As distance plays a key role in stabilising/suppressing magnetic states.[43] This is supported by studies of BaFe$_2$As$_2$, in which superconductivity can be induced by replacing Fe with Co or by applying hydrostatic pressure.[7,46,47] Although both actions results in a reduction of the unit-cell volume and the appearance of superconductivity with a similar $T_C$, the effect on various structural parameters are different or even opposite (for example the As-Fe-As angle and the Fe-As "height"). Drotziger *et al.* showed that the only parameter that might exhibit a similar behaviour with both Co substitution and pressure is the Fe-As bond length, which implies it is the decisive parameter for tuning between the SDW and superconducting ground state.[48] This is consistent with our observations which show that the magnetic transition temperature and the Fe moment can be changed just by partially replacing As with the larger isovalent Sb and without changing the carrier concentration.

In order to obtain high $T_C$ in these layered structures it is important to start from an AFM parent compound with a high $T_N$, quite a large magnetic moment[49] and the As-Fe-As angle should be close to the ideal value.[50,51] However, to achieve a superconducting state, the long range AFM order first has to be suppressed (even if a short range magnetic order still exists in the superconducting compounds and probably participate in the pairing mechanism of Cooper pairs) and at the same time the associated average magnetic moment (seen by neutron diffraction) has to be strongly reduced. With Sb substitution in LaFeAsO the As-Fe-As angle goes towards the ideal value but at the same time, the Fe-As bond increases substantially. Hence, although the structural transition temperature is reduced, the magnetic interactions are strengthened, at least up to 40% Sb and, as a consequence, superconductivity can not emerge. In the fluorine doped samples superconductivity can still exist and $T_C$ even be increased with small amounts of Sb in the structure[25,26] since fluorine doping of LaFeAsO increases the electron carrier density and weakens the magnetic interactions.[1,37] Introducing oxygen vacancies has a similar effect on the electronic structure[52] and thus it may be possible to induce superconductivity in our Sb-containing samples by making them oxygen deficient. We have preliminary results on some of our HP samples that support this idea. The improvement in the absolute value of $T_C$ in the fluorine doped compounds may be related to the reduction in the As-Fe-As angle associated with the addition of Sb. However, as the Sb content increases so does the Fe-As bond and eventually the SDW order is recovered. According to Kim *et al.*[43], an increase in the Fe-As distance of more than 0.5% from the initial value of ~2.39 Å is detrimental for inducing superconductivity and favours the magnetic ground states in $SrFe_2As_2$. This means that even though the DOS at $E_f$ is greatly enhanced in LaFeSbO (which is favourable for higher $T_C$) and the $\mu_{Fe}$ is larger as predicted by Moon *et al.*[24] and Lèbegue *et al.*[30], the large Fe-As distances means that superconductivity is unlikely to be present in pure LaFeSbO.

## V. CONCLUSIONS

In summary, we have successfully prepared polycrystalline samples of LaFeAs$_{1-x}$Sb$_x$O x≤0.4 using solid state synthesis at both ambient and high pressure. The phase diagram was explored using powder neutron diffraction, specific heat, magnetisation and resistivity measurements. Our results show that there is a decrease in both $T_S$ and $T_N$ with increased Sb content but superconductivity does not appear in the measured composition and temperature ranges. Contrary to fluorine and phosphorous substitution in LaFeAsO, increased Sb content does not cause a reduction of the magnetic moment and as a consequence superconductivity can not be induced in the fully oxygenated samples. Comparing the impact of the isoelectronic substitution of Sb and P for As on the physical and structural properties of LaFeAsO suggests that structural changes, especially the Fe-As bond length is the key parameter for controlling the strength of the magnetic interactions and the superconducting window. The Sb substituted compounds could be a good choice of family to obtain higher $T_C$ as they show a slightly higher magnetic moment and an enhanced density of states at $E_f$, if an efficient way of doping can be found. It will be interesting to further investigate the superconductivity in such partially Sb substituted arsenides.

FIG. 1. Observed (red cross), calculated (green line) and difference (pink line, below) patterns from the Rietveld analysis of (a) X-ray powder diffraction data collected at 300 K and (b) neutron powder diffraction data collected at 2 K of LaFeAs$_{0.6}$Sb$_{0.4}$O. The major peaks of the La$_2$O$_3$/La(OH)$_3$ (5%) and FeSb (6%) impurities are marked with "*".

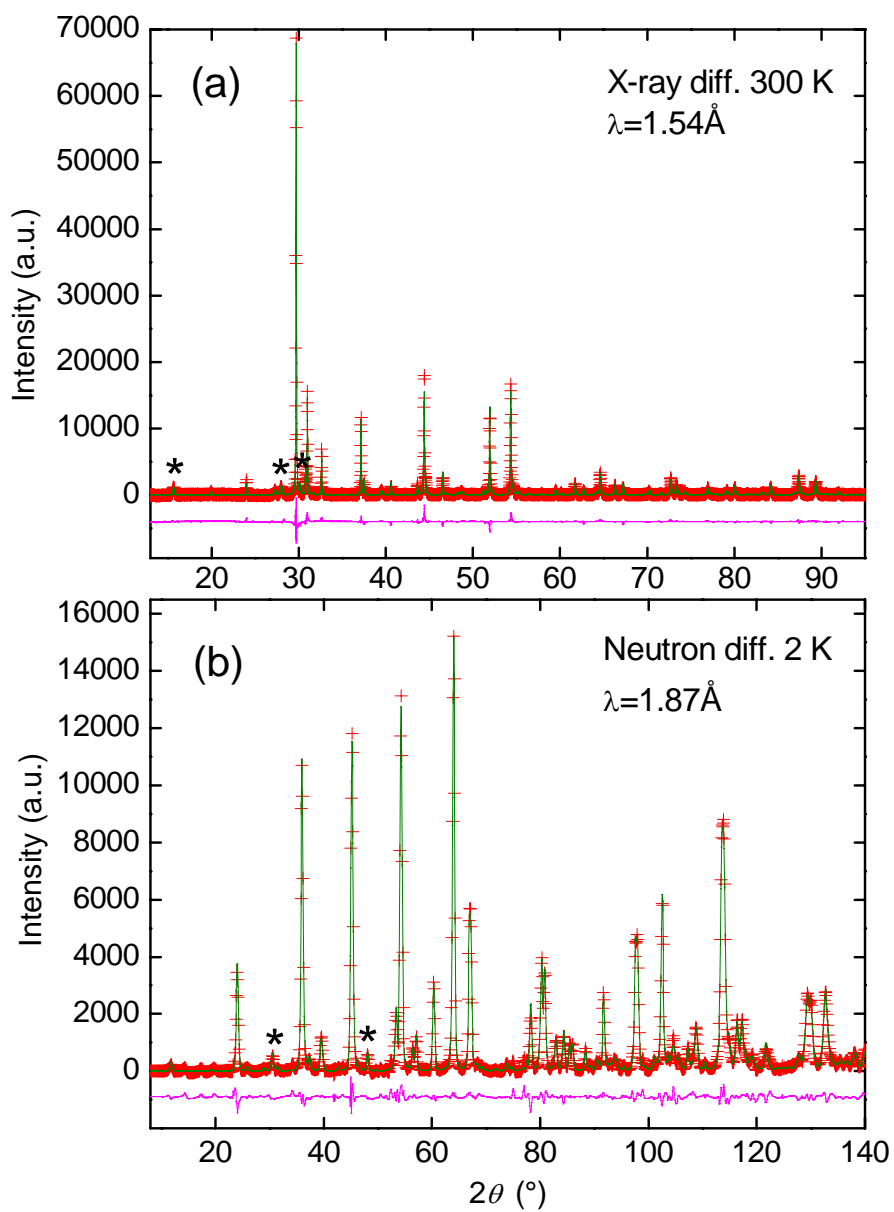

FIG. 2. Lattice parameters versus Sb content at 300 K for LaFeAs$_{1-x}$Sb$_x$O samples made with ambient (blue squares) and high (red circles) pressure synthesis. The stars represent the values predicted by DFT calculations[24,30]. Dotted lines are a guide for the eye.

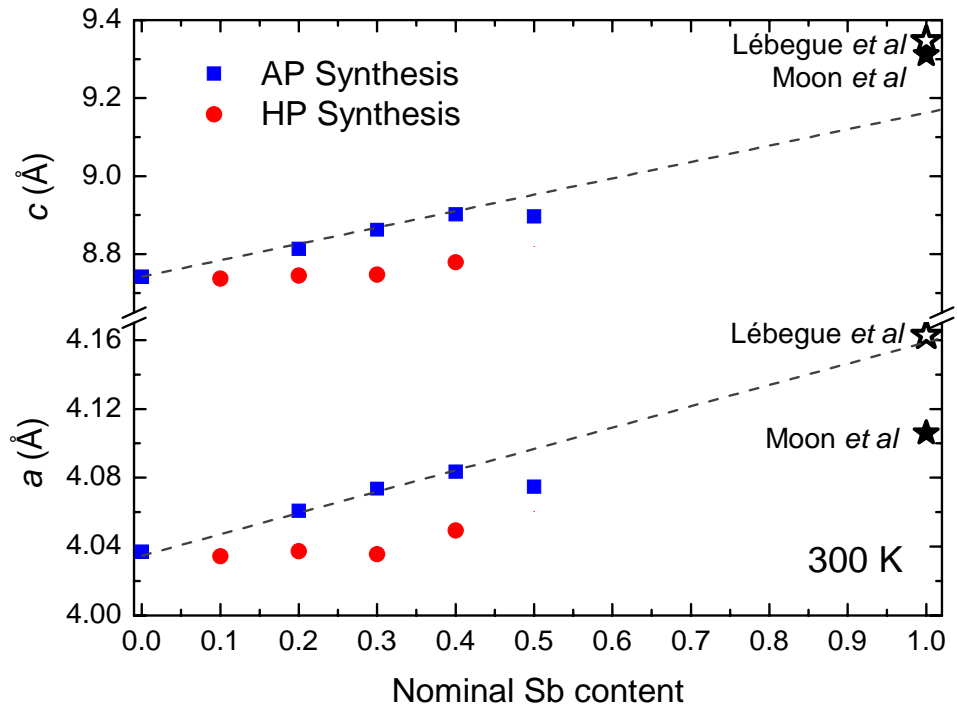

FIG. 3. Powder neutron diffraction data for LaFeAs$_{1-x}$Sb$_x$O at variable temperatures. Selected $2\theta$ range showing the change in diffraction intensities of the tetragonal (220) reflection with temperature for (a) LaFeAs$_{0.8}$Sb$_{0.2}$O and (b) LaFeAs$_{0.6}$Sb$_{0.4}$O. (c) Temperature dependence of the (1,0,3/2) and (1,2,1/2) magnetic Bragg peaks of LaFeAs$_{0.6}$Sb$_{0.4}$O

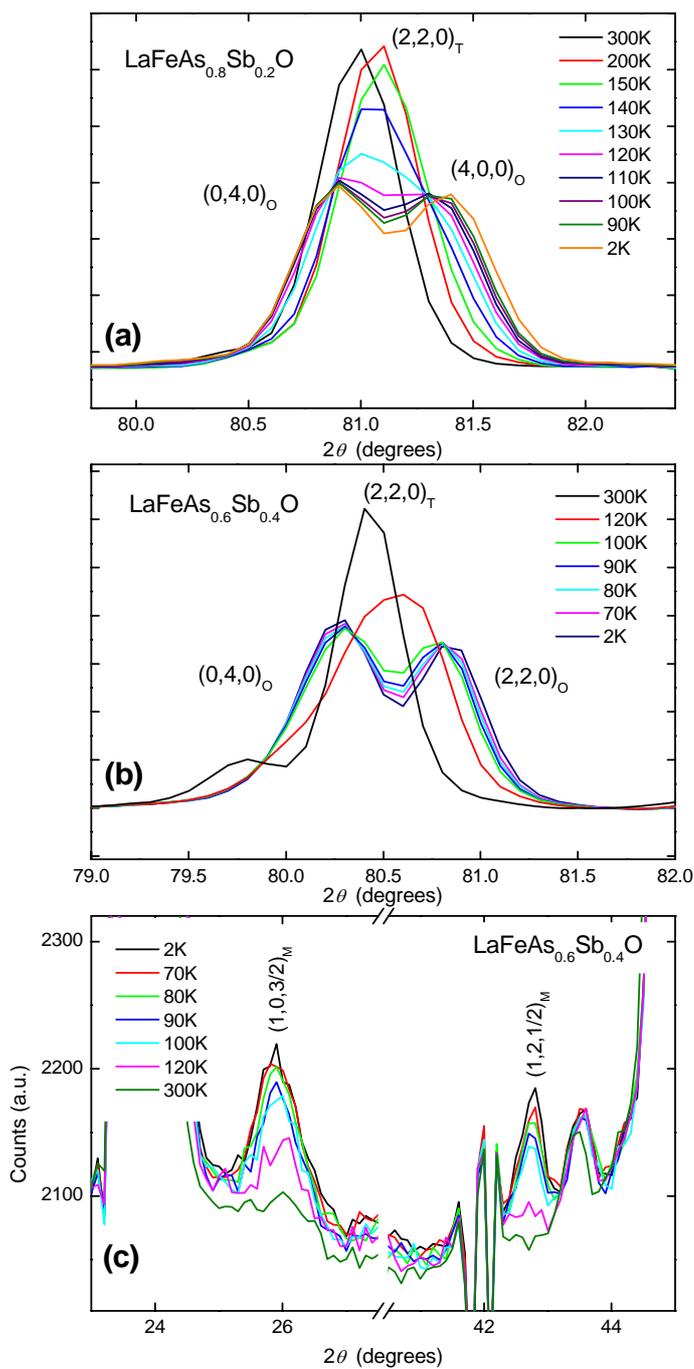

FIG. 4. Temperature dependence of (a) the unit-cell lengths for LaFeAs$_{0.8}$Sb$_{0.2}$O and LaFeAs$_{0.6}$Sb$_{0.4}$O based on neutron diffraction data (the error bars are smaller than the markers). (b) the Fe magnetic moment of LaFeAs$_{0.8}$Sb$_{0.2}$O and LaFeAs$_{0.6}$Sb$_{0.4}$O. Critical law curves are shown as a guide to the eye. The inset of (b) shows the structural phase transition expressed by the orthorhombicity parameter $(a-b)/(a+b)$. Data for x=0 are taken from ref. 32. In (a) and the inset of (b), the dashed lines are there as a guide to the eye.

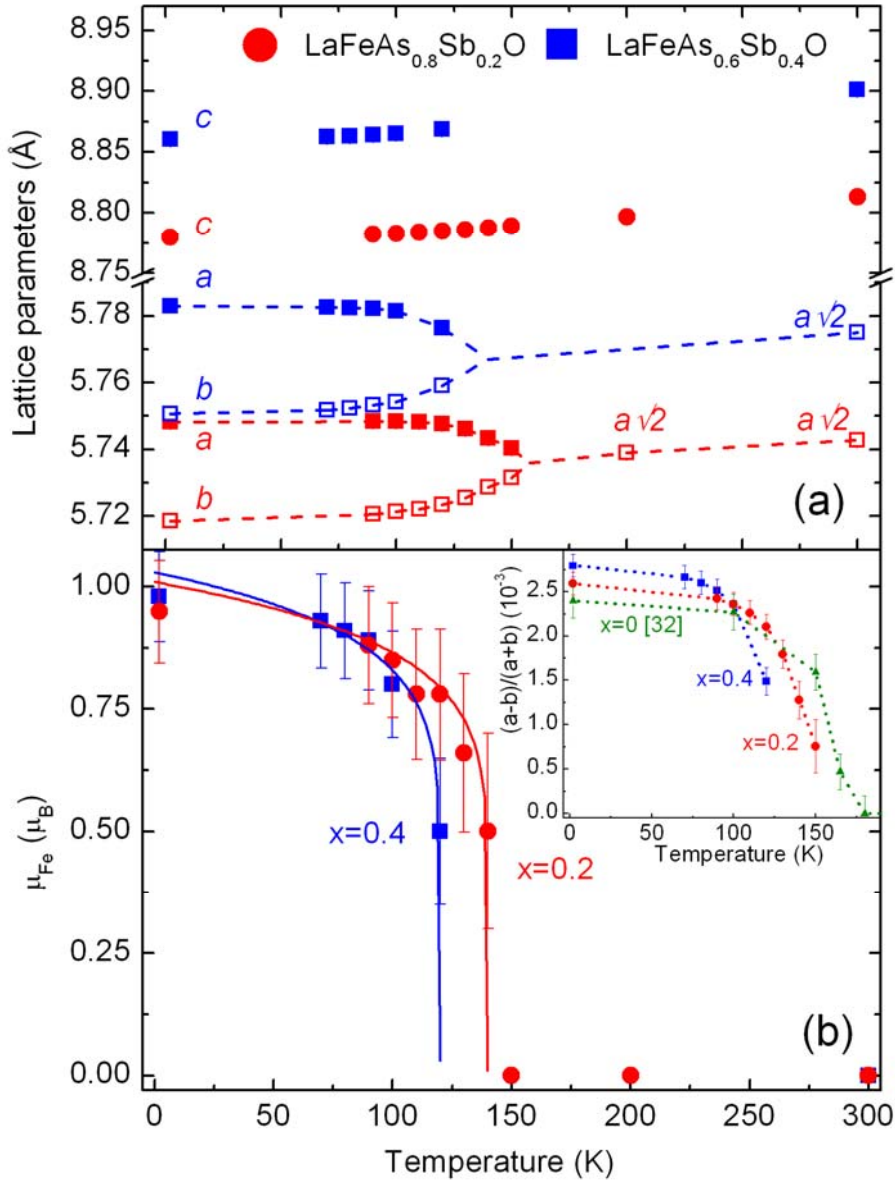

FIG. 5. Behaviour of the physical properties of LaFeAs$_{1-x}$Sb$_x$O x=0, 0.2 and 0.4 near the structural and magnetic phase transitions. (a) The anomalous contributions, $\Delta c_p/T$, to the specific heat as described in the text. The inset shows the raw data $C_p/T$. (b) Magnetisation, *M/H*, versus temperature measured at an applied field of 1 T. A Curie-Weiss type background due to the FeSb impurity has been removed from the LaFeAs$_{0.6}$Sb$_{0.4}$O data to give a clearer picture of the transition. Note that the negative values only indicate a negative deviation from the background, not a negative magnetisation. The inset shows the derivative, d(M/H)/dT. (c) Temperature dependence of zero-field resistivity ($\rho/\rho_0$). The inset shows the derivative $d\rho/dT$. The $\Delta c_p/T$, $C_p/T$, d(*M/H*)/dT and $d\rho/dT$ data sets have all been normalised by it's maximum value whereas *M/H* and $\rho$ were normalised by the value at 280 K.

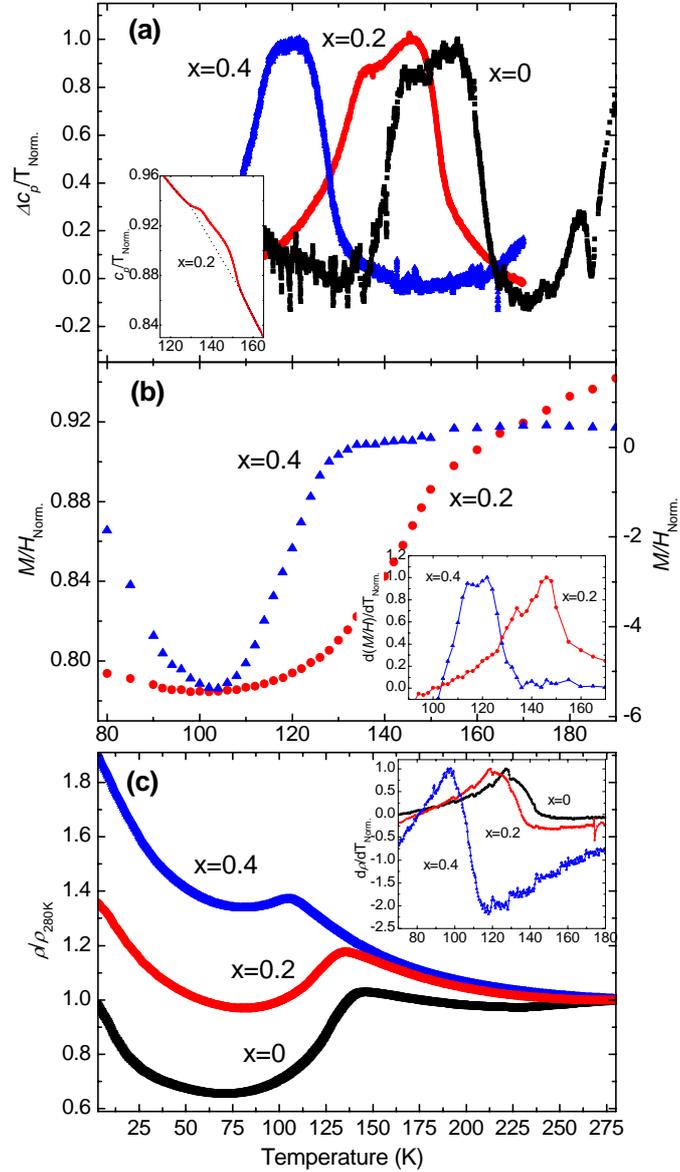

FIG. 6. (a) The variation of the structural ($T_S$) and magnetic ($T_N$) transition temperatures with Sb content estimated from resistivity (ρ), magnetisation (M/H), specific heat ($C_P$) measurements (following the approach of Klauss et al.[35] and Kondrat et al.[36]) and neutron diffraction data (NPD). $T_S$ is represented by open symbols while $T_N$ is indicated by full ones. (b) Evolution of the structural parameters of LaFeAs$_{1-x}$Sb$_x$O as a function of Sb doping (positive x) obtained from the Rietveld fits of the 300K neutron diffraction data. Negative values of x indicate P doping and the corresponding structural parameters were taken from Wang et al.[9]. For comparison the parameters for x=1.0 from Lébegue et al.[30] are also plotted. Dotted lines are added as a guide to the eye. The data was normalised by the values for x=0.

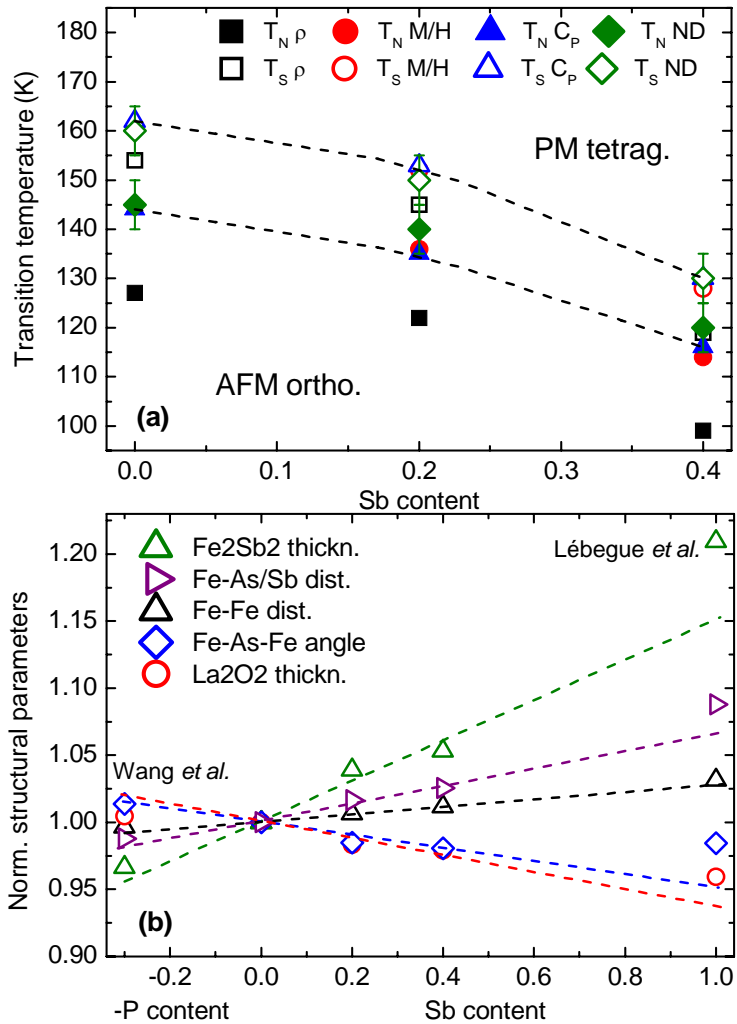

TABLE I. Crystallographic data, magnetic moment and structural and magnetic transition temperatures of LaFeAs$_{1-x}$Sb$_x$O (x = 0.2 and 0.4) from the Rietveld refinements of the D20 neutron diffraction patterns collected at 300K and at 2K. The values for x = 0 taken from ref. 32 has been added for comparison.

| | 300 K, space group P4/nmm | | | 2 K, space group cmme | | |
|---|---|---|---|---|---|---|
| | x=0[a] | x=0.2 | x=0.4 | x=0[a] | x=0.2 | x=0.4 |
| a (Å) | 4.0322(2) | 4.06076(6) | 4.08354(5) | 5.7063(4) | 5.7482(1) | 5.7830(1) |
| b (Å) | 4.0322(2) | 4.06076(6) | 4.08354(5) | 5.6788(4) | 5.7185(1) | 5.7507(1) |
| c (Å) | 8.7364(4) | 8.8133(2) | 8.9014(2) | 8.7094(6) | 8.7799(2) | 8.8606(2) |
| V (Å$^3$) | 142.336(4) | 145.330(6) | 148.434(5) | 282.23(2) | 288.62(1) | 294.67(1) |
| z La | 0.1416(4) | 0.1380(3) | 0.1363(2) | 0.1420(4) | 0.1383(2) | 0.1375(2) |
| z As/Sb | 0.6508(5) | 0.6558(4) | 0.6568(3) | 0.6505(5) | 0.6561(3) | 0.6557(3) |
| La-O (Å) | 2.365(5) | 2.367(1) | 2.374(1) | 2.362(5) | 2.363(1) | 2.375(1) |
| Fe-As/Sb (Å) | 2.408(5) | 2.451(2) | 2.473(2) | 2.402(5) | 2.447(2) | 2.462(2) |
| As-Fe-As (°) | 113.7(1) | 111.9(1) | 111.33(9) | 113.8(1) | 112.4 (1) | 111.8 (1) |
| Rp, wRp (%), $\chi^2$ | c | 1.73, 2.44, 2.0 | 2.15, 2.94, 2.31 | c | 1.58, 2.16, 1.61 | 2.15, 2.98, 2.44 |
| μ$_{Fe}$ (μ$_B$) | n.a | n.a | n.a | 0.63(1) | 0.95(18) | 0.98(13) |
| T$_S$, T$_N$ (K) | 160, 145 [b] | 150,140 (±5) | 130, 120 (±5) | n.a | n.a | n.a |

[a] Reference [32]
[b] Reference [31]
[c] Not available

**Acknowledgements:**

This work was supported by the French National Research Agency, Grant No. ANR-09-Blanc-0211 SupraTetrafer. We are grateful to Pierre Strobel (Institut Néel, Grenoble, France) for his help with the tube sealing technique and for useful discussions.